\documentclass[]{aa}  

\usepackage{graphicx}
\usepackage{txfonts}
\usepackage[colorlinks]{hyperref}
\hypersetup{
     colorlinks=true,
     linkcolor=blue,
     filecolor=blue,
     citecolor =blue,      
     urlcolor=blue,
     }
     
\usepackage{graphicx}	
\usepackage{amsmath}	
\usepackage{amssymb}	

\newcommand\gaia{\textit{Gaia}}

\newcommand\gdrthree{\gaia~DR3}
\newcommand\gedrthree{\gaia~EDR3}

\begin{document}

  \title{Galactic spiral structure revealed by \gedrthree\  }
   
   \titlerunning{Galactic spiral structure revealed by \gedrthree\  }



  \author{E. Poggio\inst{1,2} \and R. Drimmel\inst{2} \and T. Cantat-Gaudin\inst{3}\and P. Ramos\inst{4} \and V. Ripepi\inst{5} \and E. Zari\inst{6} \and  R. Andrae\inst{6} \and R. Blomme\inst{7} \and\\ L. Chemin\inst{8} \and 
  G. Clementini\inst{9} \and F. Figueras\inst{3} \and M. Fouesneau\inst{6}  \and Y. Frémat\inst{7} \and A. Lobel\inst{7} \and  \\ D. J. Marshall\inst{10,11} \and  T. Muraveva\inst{9} \and M. Romero-G\'omez\inst{3} }
 

  \institute{Université Côte d’Azur, Observatoire de la Côte d’Azur, CNRS, Laboratoire Lagrange, France\\
              \email{poggio.eloisa@gmail.com}
         \and
             Osservatorio Astrofisico di Torino, Istituto Nazionale di Astrofisica (INAF), I-10025 Pino Torinese, Italy\\
             \email{ronald.drimmel@inaf.it}
        \and
        Institut de Ci\`encies del Cosmos, Universitat de Barcelona (IEEC-UB), Mart\'i i Franqu\`es 1, E-08028 Barcelona, Spain\label{ICC}
        \and
        Observatoire astronomique de Strasbourg, Universit{\'e} de Strasbourg, CNRS, 11 rue de l’Universit{\'e}, 67000 Strasbourg, France
        \and
        INAF-Osservatorio Astronomico di Capodimonte, Salita Moiariello 16,
        I-80131, Naples, Italy
        \and
        Max-Planck-Institut f\"ur Astronomie, K\"onigstuhl 17 D-69117 Heidelberg, Germany
        \and
        Royal Observatory of Belgium, Ringlaan 3, B-1180 Brussels, Belgium
        \and
        Centro de Astronom\'ia, Universidad de Antofagasta, Avda. U. de Antofagasta 02800, Antofagasta, Chile
        \and
        INAF-Osservatorio di Astrofisica e Scienza dello Spazio di Bologna, Via Gobetti 93/3, I - 40129 Bologna, Italy
        \and
        Université de Toulouse, UPS-OMP, IRAP, F-31028 Toulouse cedex 4, France
        \and
        CNRS, IRAP, 9 Av. colonel Roche, BP 44346, F-31028 Toulouse cedex 4, France
             }

   \date{Received XXXX; accepted YYY}

  \abstract{ Using the astrometry and integrated photometry from the \gaia\ Early Data Release 3 (EDR3), we map the density variations in the distribution of young Upper Main Sequence (UMS) stars, open clusters and classical Cepheids in the Galactic disk within several kiloparsecs of the Sun.
  Maps of relative over/under-dense regions for UMS stars in the Galactic disk are derived 
  using both bivariate kernel density estimators and wavelet transformations.
  The resulting overdensity maps exhibit large-scale arches, that extend in a clumpy but coherent way over the entire sampled volume, 
  indicating the location of the spiral arms segments in the vicinity of the Sun. Peaks in the UMS overdensity are well-matched by the distribution of young and intrinsically bright open clusters. By applying a wavelet transformation to a sample of classical Cepheids, we find that their overdensities possibly extend the spiral arm segments on a larger scale ($\simeq 10$ kpc from the Sun). While the resulting map based on the UMS sample is generally consistent with previous models of the Sagittarius-Carina spiral arm, the geometry of the arms in the III quadrant (galactic longitudes $180^\circ < l < 270^\circ$) differs significantly from many previous models.
  In particular we find that our maps favour a larger pitch angle for the Perseus arm, and that the Local Arm extends into the III quadrant at least 4 kpc past the Sun’s position, giving it a total length of at least 8 kpc.
 
 }





   \maketitle
%

\section{Introduction}

The first indication of the large-scale structure of the Milky Way came with the realization that it was but one of a large class of galaxies, of which our neighbor Andromeda served as the first archetype. At that point our collective mental picture of the Milky Way was that of a spiral galaxy, though we remained ignorant of the number and position of the spiral arms.
The first hint of spiral structure in the vicinity of the Sun was found in the distribution of short-lived high mass stars and early 21 cm surveys \citep{Morgan:1952,Morgan:1953,Christiansen:1952}, showing three spiral arm segments. The transparency of the interstellar medium to radio wavelengths soon allowed a confirmation that the distribution of gas in the disk had spiral arms on large scales \citep{VanDeHulst:1954,Kerr:1957,Oort:1958}, though a clear picture remained lacking. 

The large-scale spiral structure would become evident after considerable effort by combining both radio and optical data of HII regions, with the definitive work of \citet{Georgelin:1976}. They convincingly traced out four spiral arms, and their mapping would later be improved and extended by \citet{Taylor:1993}, \citet{Bland-Hawthorn:2002} and \citealt{Russeil2003}.
Advances in 21 cm surveys and modelling also showed four arms in the outer galaxy, where kinematic distances are unambiguous, out to about 25 kpc from the Galactic center \citep{Levine:2006}. More recently \citet{Reid:2014,Reid:2019} has mapped the spiral arms using absolute radio astrometry of maser sources, thought to be powered by young stellar objects. The spiral structure traced by the masers is again broadly consistent with the picture painted by the HII mapping of a 4-armed spiral. Classical Cepheids have also begun to be used to trace the spiral arms on larger scales \citep{Skowron:2019,Veselova:2020}.
While these and most other models agree on the number of major arms in the Milky Way, as traced by star formation products, there are some differences in their position, shape and pitch angles. The most marked difference is between the spiral arms in the outer disk of \citet{Levine:2006}, with a pitch angle of approximately $24^\circ$, and other models based on tracers in the inner Galaxy, which typically have pitch angles between 12 and 13 degrees \citep{Vallee:2005}. In particular, these models differ in their proposed tracing of the Perseus arm. 

In most spiral arm models the Local Arm, one of the three spiral arm segments detected and passing in close vicinity of the Sun, is either not present or added as a small local feature, a relatively minor spur or arm segment. Nevertheless, due to its proximity, this feature dominates the observed distribution on the sky of bright young stars \citep{Poggio:2017}. 
Only recently, with the mapping of radio masers associated with young stellar objects, has it become evident that the Local Arm extends further into the first quadrant (galactic longitudes $0^\circ > l > 90^\circ$) than previously thought \citep{Xu:2016}. \cite{Miyachi:2019} found that the stellar Local Arm has a larger pitch angle compared to the one based on High-Mass Star Forming Regions from \citet{Xu:2016}.  


For recent reviews of our current knowledge of the spiral structure of the Milky Way, see \citet{XuRev:2018} and \citet{Shen:2020}. 

With the advent of the \gaia\ mission \citep{GaiaColl:2016,GaiaColl:2018,GaiaCollEDR3:2020} and its billion plus stellar parallaxes, there has been a resurgence in mapping the spiral structure in the vicinity of the Sun, as seen in young populations.  \citet{CantatGaudin:2018,CantatGaudin:2020} and \citet{Kounkel:2020} have identified and mapped open clusters, while \citet{Xu:2018}, \citet{Chen:2019} and \citet{Xu:2021} have investigated the distribution of OB stars as well as relevant maser sources with radio astrometry. Combining \gaia\ astrometry with radial velocity data, evidence of spiral arm structure is also found in the kinematics of stars in the form of non-circular streaming motions \citep{GaiaCollab:2018, Eilers:2020, Khoperskov:2020}. 



In this contribution we investigate the spiral arm structure, as seen in the surface density of three types of young stellar tracers: upper main sequence stars, open clusters and classical Cepheids.
The paper is structured as follows: in Section \ref{Sec:Data}, we describe the datasets used in this work; the methods and results are presented in Section \ref{Sec:Results}; we discuss our results in Section \ref{Sec:DiscConcl} and present our conclusions in Section \ref{Sec:concl}.



\section{Data} \label{Sec:Data}
In the following subsections, we give an overview of the datasets used in the present work.

\subsection{Upper Main Sequence stars}
Here we use two samples of Upper Main Sequence (UMS) stars:
\begin{itemize}
    \item \emph{The "P18" UMS sample}, which contains the UMS stars from \cite[][hereafter P18]{Poggio:2018}, but 
    with \gedrthree\  astrometry \citep{GaiaCollEDR3:2020}. In P18, the selection of the stars was based on two stages. First, using photometric measurements from the 2-Micron All Sky Survey \citep[2MASS,][]{Skrutskie:2006} and \emph{Gaia} \citep{Evans:2018}, a preliminary selection was performed, selecting stars with $G < 15.5$ and colors consistent with instrinsically blue sources subject to reddening. The selection was further refined via a probabilistic approach based on \gaia\ astrometry, with the aim of selecting stars that are likely to be brighter than a given (reddened) absolute magnitude $M_{lim}$, chosen to be roughly consistent with a B3V type star. For this sample, distances are recalculated using the bayesian procedure described in P18, but using \gedrthree\ astrometry, corrected for the mean zero point parallax offset following \citet{Lindegren2020}, resulting in 603\,787 selected sources. 
    
    Due to quality cuts on the 2MASS photometry it was found that many nearby (bright) blue sources were missing. We therefore selected \gedrthree\ sources with apparent magnitude $G< 15.5$, color $(G_{BP} - G_{RP}) < 0$, and parallax $\varpi > 0.2$ mas. Additionally, we only select stars satisfying the relation $G + 5 \log(\varpi/1000) + 5 < 0$, aimed at removing white dwarf contaminants. 
    This procedure gave us 6827 stars, of which 3259 are not in the P18 sample described above. Of these, only 50 were found to have a parallax signal-to-noise smaller than 5, and we excluded them from our sample.
    For the remaining 3209 stars we simply invert the parallax to obtain their distance. In any case, as will be discussed below, the role played by these 3209 nearby UMS stars is almost irrelevant to the results presented in this contribution, as they only cause modest changes within $0.3 - 0.4$ kpc from the Sun, while this work aims to map the spiral structure of the Galaxy on a larger scale. 
    
    The final UMS P18 sample contains {\bf $606 \, 219$} stars, after removing duplicated sources resulting from the crossmatch between \emph{Gaia} DR2 and DR3.
    
    \item \emph{The "Z21" UMS sample}, which contains the stars in the 'filtered' sample from \cite{Zari:2021}.
    Similarly to the P18, Z21 selected their UMS initial sample by applying simple color cuts combining \textit{Gaia} DR3 and 2MASS photometry. To select intrinsically luminous stars (of spectral type earlier than $\sim$B7V), they restricted their sample to stars with absolute magnitude in the 2MASS $K_s$ band of $M_{K_s} < 0 \, \mathrm{mag}$.  Z21 refer to such sample as the 'target sample', as it is devised for spectroscopic follow-up. 
    Finally, Z21 cleaned their target sample of sources with spurious astrometric solutions, using the classifier presented in \cite{Rybizki:2021}, and for sources that do not follow disk kinematics of a young stellar population. Their 'filtered' sample consists of 435 273 stars. Z21 estimated distances by using a model designed to  reproduce the properties of their  data-set in terms of its spatial and luminosity distribution, and the additional information that stars  belonging to their sample should follow Galactic rotation, with a small, typical velocity dispersion. They refer to such distances as 'astro-kinematic' distances.

\end{itemize}

As anticipated above, the two samples are expected to trace a similar stellar population. As a confirmation, we found that they have $265 \, 004$ sources in common. 
Differences between the P18 and Z21 samples are presumably due to the cuts applied on photometry, which might exclude numerous late B-type stars in the UMS P18 sample, and astrometric quality cuts applied in the UMS Z21 sample. The results presented in Section \ref{Sec:Results} have been tested using both UMS samples (and their corresponding different distance estimates to the stars), in order to verify the robustness of our findings.


\subsection{Open Clusters} \label{Sec:OCs}
We use the sample of open clusters with members and ages published by \citet{CantatGaudin:2020}, which was based on \textit{Gaia}~DR2 data. All of their member stars are brighter than $G$=18 and 98.4\% of them have an EDR3 counterpart with the same source id number, which we use to recover their EDR3 astrometry. 
In order to refine the mean astrometry of the clusters, including their mean parallax, we reject as members those stars whose EDR3 proper motion is discrepant from the cluster median by more than 3-$\sigma$, and recompute the median EDR3 parallax. 

Since the uncertainty on the median parallax of a cluster is much smaller than for individual stars, we estimate cluster distances by inverting their median EDR3 parallax, after correcting for the known negative parallax zero point of -17$\mu$as.

For this study we select the 687 open clusters younger than 100\,Myr. Furthermore, we refine the selection by only considering the 353 open clusters with more than 5 members brighter than absolute magnitude $M_G = 0$ (computed taking into account their distance and interstellar extinction). This additional cut has been applied to select intrinsically bright clusters, and compensate for observational biases, which otherwise would cause an over-sampling of clusters in the solar neighborhood. In the following, we will refer to this sample as the young and intrinsically bright open cluster dataset. 

The complete open cluster sample can be downloaded from CDS. 



\begin{figure*}
   \centering
    \includegraphics[width=1.0\textwidth]{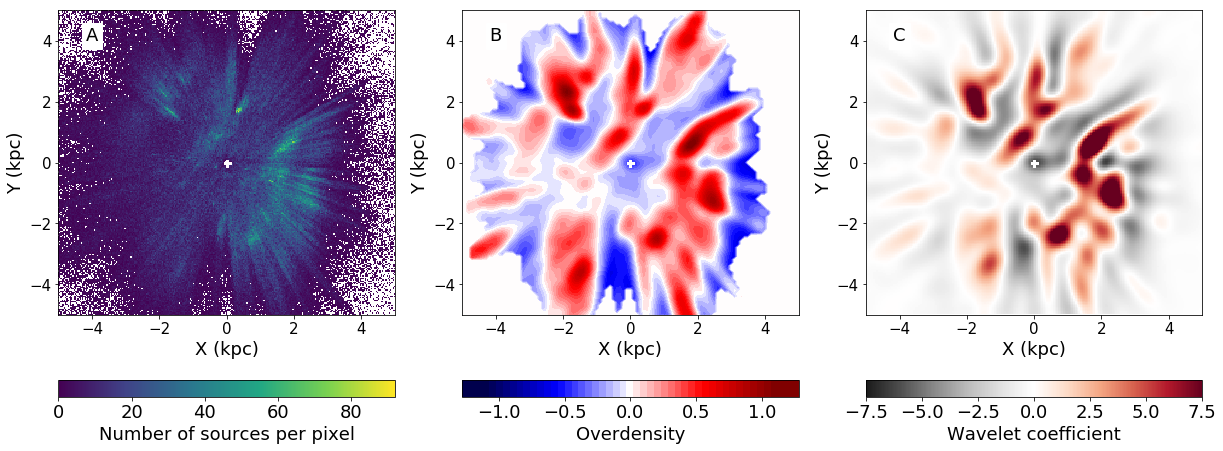}
   \caption{\emph{Panel A:} Face-on view of the UMS P18 dataset in the Galactic disk. The position of the Sun is shown by the white cross in (X,Y)=(0,0). The Galactic center is to the right, in (X,Y)=($R_{\odot}$,0), and the Galaxy is rotating clockwise. \emph{Panel B:} Same as Panel A, but showing the measured overdensity using the UMS P18 dataset, based on a local density scale length $0.3$ kpc. Only points with $\Sigma(x,y)>$ 0.003 are plotted, in order to remove regions where the statistics is too low.
   The corresponding plot for the UMS Z21 sample can be found in Figure \ref{Fig:Z21_UMS_overdens_map}.
   \emph{Panel C:} Same as Panel A, but showing the wavelet transformation at the scale 3 (size$\sim$0.4 kpc). A different version of Panel B and C using a larger scale length can be found in Figure \ref{Fig:h08} (see Appendix). The maps shown in panel B and C are publicly available at \url{https://github.com/epoggio/Spiral_arms_EDR3.git.}}
\label{Fig:overview}%
\end{figure*}

\subsection{Classical Cepheids} \label{Sec:Cepheids}

We compiled a list of classical Cepheids (DCEPs) taken from a variety of sources. In more detail, we adopted the list published by \gaia\ DR2 \citep{Clementini:2019} as revised by \citet{Ripepi:2019}, the compilation by \citet{Skowron:2019}, the recent new discoveries  by the OGLE \citep[Optical Gravitational Lensing Experiment][]{Udalski:2018,Sos:2020} and  ZTF \citep[Zwicky Transient Facility,][]{Chen:2020} surveys.  
Merging the above catalogs, we obtained a sample of 2004 and 873 fundamental (DCEP\_F) and first overtone (DCEP\_1O) DCEPs, respectively. Note that the sample includes 238 mixed-mode pulsators that we assigned to DCEP\_F or DCEP\_1O on the basis of the dominant mode (85 and 153 DCEPs with dominant fundamental and 1O mode, respectively)\footnote{The details about the construction of the sample will be given elsewhere (Ripepi et al. in preparation)}. 
The literature sample was cross-matched with EDR3 to obtain parallaxes and  $G,G_{BP},G_{RP}$ magnitudes\footnote{The correct average magnitude of a DCEP is normally calculated by integrating the light curve in intensity and then transformed back into magnitude. The \gaia\ magnitudes are arithmetic and can differ from the "correct" ones by several hundredths of magnitude \citep[see e.g.][]{Caputo:1999}. However, using DR2 results we estimate that for about 80\% of the DCEPs the arithmetic mean of the Wesenheit magnitude defined in the text differs by less than 0.02 mag from the intensity-averaged one. In our case, the impact of such an uncertainty on the distance is generally negligible with respect to the uncertainty introduced by the parallaxes.} \citep{Riello:2020}.  

To derive the distances for our DCEP sample, we first calculated the individual Wesenheit magnitudes, that are particularly useful as they are reddening-free by definition \citep[][]{Madore:1982}. The formulation for the apparent Wesenheit magnitude in the \gaia\ bands is $w=G-1.90*(G_{BP}-G_{RP})$ \citep{Ripepi:2019}. Assuming the periods and modes from the literature, the individual DCEPs distances (in pc) can be calculated directly from the distance modulus definition: $w-W=-5+5*\log d$(pc), where the absolute Wesenheit magnitude W is estimated through the PW relation. PW relations for both DCEP\_F and DCEP\_1O have been published by \citet{Ripepi:2019} based on \gaia\ DR2 parallaxes. However, the advent of the more precise EDR3 parallaxes led us to recalculate these relations, as described in Appendix \ref{appendix_PWrelation}.
Finally, the DCEP individual distances for the enlarged sample of 2004 DCEP\_F and 873 DCEP\_1O DCEPs were then calculated directly from the definition of distance modulus. 

\begin{figure*}
   \centering
   \includegraphics[width=1.0\textwidth]{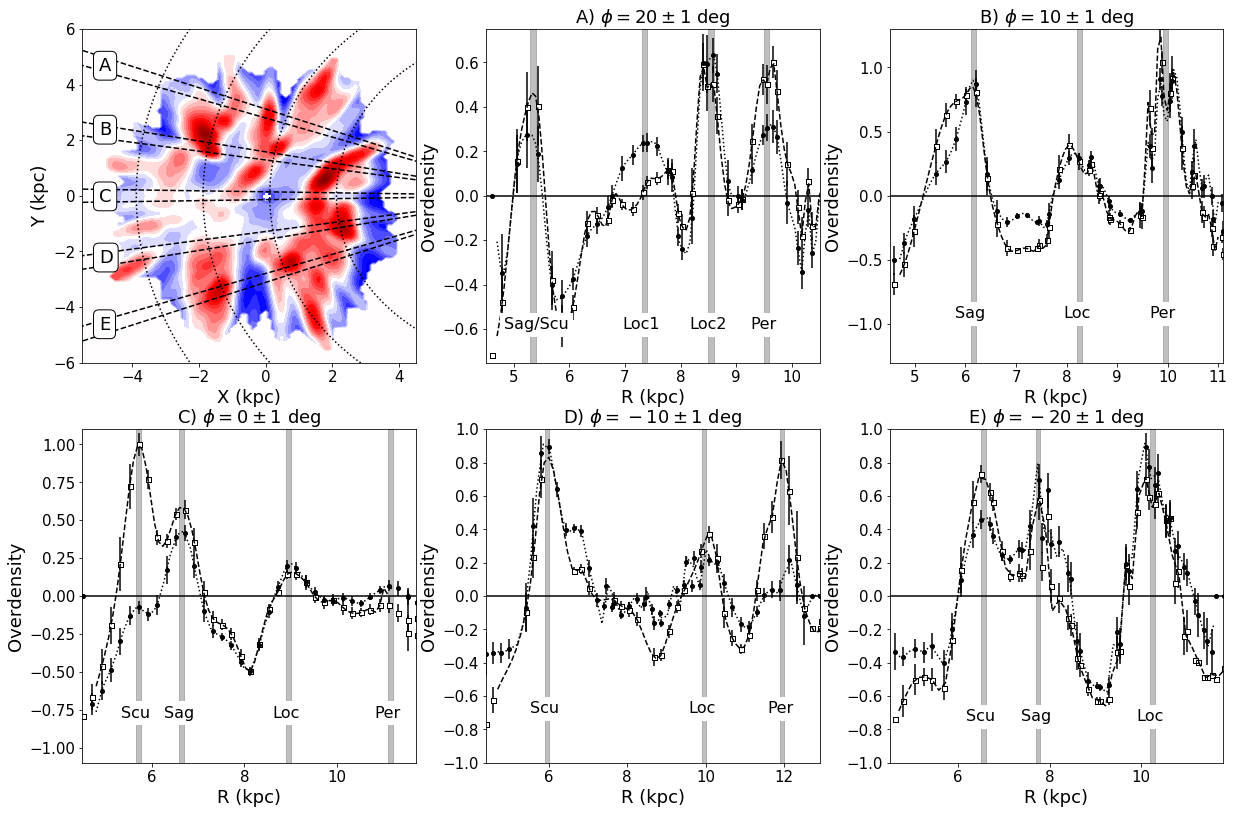}
   \caption{\emph{Top left panel:} Same as Figure \ref{Fig:overview}B, but with some geometric references superimposed. Dotted lines from left to right show the points with constant Galactocentric radius R=12,10,8,6 kpc, respectively. Dashed lines show the five selected slices (A,B,C,D,E), which are separated by 10 deg in Galactic azimuth $\phi$ and of 2 deg in width. \emph{Panels A to E:} The profile of the measured overdensity as a function of R for the UMS P18 sample (black dots) and the Z21 sample (white squares). Error bars show bootstrap uncertainties, calculated as explained in Appendix \ref{appendix_uncertainties}. For computational reasons, black dots/white squares and relative errorbars were calculated for each slice and then connected using splines, as shown by the dotted (dashed) lines, to give a visual impression of how the overdensity varies at different radii. Each peak of the measured overdensity has been identified with a spiral arm in the Milky Way, as indicated by the vertical grey lines and the corresponding labels. The identification of each peak is not due to an assumed specific model, but is simply based on the geometric appearance of the top left panel (see text). }
              \label{Fig:structure}%
\end{figure*}

In the context of this paper, it is important to have a measure of the DCEPs age. These variables obey period-age (PA) and period-age-color (PAC) relations \citep[see e.g.][]{Bono:2005}. In this work we estimated the ages using the updated PA relation by \citet{Desomma:2020}\footnote{We preferred the PA over the PAC, as to use colors we would need to know the individual reddening values, which are rather uncertain for disk objects such as DCEPs}. In particular, we adopted the PA relations for models without overshooting as this allows us to use both DCEP\_F and DCEP\_1O variables \citep[see][for details]{Desomma:2020}. 
Using the PA relations we select 1923 Cepheids with ages younger than $10^8$ years. These DCEPs have ages consistent with the young and intrinsically bright open clusters defined in Sect.~\ref{Sec:OCs}.


\section{Results} \label{Sec:Results}
Figure \ref{Fig:overview} gives an overview of the UMS P18 dataset. The spatial distribution of the stars in the X-Y plane of the Galactic disk is presented in Figure \ref{Fig:overview}A, where the Sun is located at (X,Y)=(0,0) kpc. The lack of stars near the Sun's position ($\lessapprox 0.2$ kpc) is due to our selection criteria, as mentioned in the previous section. The spatial distribution exhibits some dense clumps, which correspond to well-known OB associations, i.e. Cygnus, Carina, Cassiopeia and Vela, amongst others.
Also to be noted are radial features with respect to the Sun's position that become increasingly evident at heliocentric distances greater than about 2.5 kpc. These are artificial features \emph{not} due to distance uncertainties, but rather are "shadow cones" produced from foreground extinction in concert with the magnitude limit of this sample. For example, a prominent shadow cone is evident starting just beyond a dense clump of young stars at about $(X,Y) = (0,2)$ kpc (Cygnus). In any case, any features showing a clear (radial or otherwise) symmetry with respect to the Sun's position should be treated with caution. 

To make the spiral structure more evident, we map the stellar \emph{overdensity} $\Delta_{\Sigma}$, defined as
\begin{equation}
    \Delta_{\Sigma} (X,Y) = \frac{ \Sigma (X,Y) \, -  \langle \, \Sigma (X,Y) \, \rangle }{\langle \, \Sigma (X,Y) \, \rangle } = \frac{ \Sigma (X,Y) }{\langle \, \Sigma (X,Y) \, \rangle } - 1
    \quad,
    \label{Eq:overdensity}
\end{equation}
where $\Sigma(X,Y)$ is the local density at the position $(X,Y)$ in the Galactic plane, and $\langle \, \Sigma (X,Y) \, \rangle$ is the mean density. Both $\Sigma(X,Y)$ and $\langle \, \Sigma (X,Y) \, \rangle$ are calculated via a bivariate kernel density estimator, but using two different bandwidths, i.e. 0.3 kpc and 2 kpc, respectively, for the local and mean density. Details on the calculation of $\Sigma(X,Y)$ and $\langle \, \Sigma (X,Y) \, \rangle$ are given in Appendix \ref{appendix_method}. 
The overdensity measured from the UMS P18 dataset is shown in Figure \ref{Fig:overview}B. The corresponding plot using the UMS Z21 sample can be found in Appendix (Figure \ref{Fig:Z21_UMS_overdens_map}). Both maps are very similar, though the UMS Z21 sample shows more clearly the inner Scutum arm, while the UMS P18 sample seems to have a better sampling of the Perseus arm.

To test the robustness of the obtained map, we also apply an alternative method using a wavelet transform to map overdense regions. (Details are given in Appendix \ref{appendix_wavelet}.)
Figure \ref{Fig:overview}C shows the obtained map, which is consistent with Figure \ref{Fig:overview}B. 

The global picture that emerges from Figure \ref{Fig:overview} is that the distribution of the UMS stars is far from axisymmetric. The stars preferentially reside along arch-like features, which are hardly visible in the raw spatial distribution (Figure \ref{Fig:overview}A), but become quite evident when the over/under-dense regions in the Galactic disk are mapped (Figure \ref{Fig:overview}B and \ref{Fig:overview}C). These features in the spatial distribution of stars can be interpreted as coinciding with spiral arm segments in the vicinity of the Sun.
Specifically, we discern three inclined red stripes in Figure \ref{Fig:overview}B and \ref{Fig:overview}C, corresponding (from left to right) to the Perseus arm, the local arm nearest the Sun, and an inner band to the Sagittarius-Carina and Scutum arms, respectively.




In Figure \ref{Fig:structure}, we dissect Figure \ref{Fig:overview}B into slices along the Galactocentric radius R, spaced by 10 $\deg$ in Galactic azimuth. The assumed distance to the Galactic center is $R_{\odot} = 8.122$ kpc \citep{Gravity:2018}. Each slice illustrates the measured overdensity features in the form of peaks/bumps as a function of R. We also compare the results obtained with the UMS P18 sample to the overdensities derived from the UMS Z21 dataset. Notwithstanding the different selection criteria used to define the two UMS samples, and the different methods to calculate distances, Figure \ref{Fig:structure} shows that most of the overdensity peaks are located at similar radii. 

For a given slice, each peak in the overdensity can be typically identified with a spiral arm of the Milky Way. By comparing the XY-map configuration (Figure \ref{Fig:structure}, top left) to the five slices, we can map the spiral arms locii throughout the Galactic plane. 

The most external spiral-arm segment of the overdensity map, identified as the Perseus arm, is located at $R \sim 10$ kpc at $\phi \sim 20 \deg$, and extends in a clumpy but coherent way until $R \sim 12$ kpc, $\phi \sim -10 \deg$. Much of this clumpiness is an artifact of foreground extinction, as evidenced by their radial orientation with respect to the Sun.  At smaller Galactocentric radii, we can find an overdensity arch-like feature that passes near the Sun, and extends from $R \sim 8$ kpc at $\phi \sim 20 \deg$ to $R \sim 10$ kpc at $\phi \sim -20 \deg$. We identify this feature with the Local Arm. We also note that, at $\phi \sim 20 \deg$, the Local Arm exhibits a bifurcation, which might be an artifact caused by foreground extinction, as discussed above. In the innermost regions, we can find the Sagittarius arm, extending from $R \sim 6$ kpc at $\phi \sim 10 \deg$ to $R \sim 8$ kpc at $\phi \sim -20 \deg$. It is not clear whether the overdensity found at $R \sim 5-5.5$ kpc and $\phi \sim 20 \deg$ corresponds to the Sagittarius or Scutum arm, which is the innermost spiral arm explored in this work. Scutum extends from $R \sim 5.5$ kpc at $\phi \sim 0 \deg$ to $R \sim 8$ kpc at $\phi \sim -30 \deg$.

In Figure \ref{Fig:structure}, uncertainties are calculated as described in Appendix \ref{appendix_uncertainties}. As we can see, the statistical significance of the detected over-/under-densities at a given point strongly depends on the position in the Galactic plane. However, in this work, rather than focusing on the signal-to-noise ratio for each single point, we are more interested in understanding whether the observed global structures (i.e. the arch-like features mentioned above) are statistically significant. To this end, we construct a grid in the $(X,Y)$ plane, where points are spaced by 200 pc along each coordinate. We then select the points of the grid lying along three spiral features identified in the overdensity map, that is (from left to right): the Perseus arm, the Local Arm, and the final one, which includes both the Sagittarius-Carina and Scutum arms (here considered together since they are difficult to clearly separate). We then construct wide bins of 10 $\deg$ in Galactic azimuth $\phi$ for each spiral feature over its entire extent. Afterwards, for each feature, we check that the overdensity signal contained in all Galactic azimuth bins is significant, that is, it contains points with a signal-to-noise ratio greater than 3. Of course, as can be noted, the features are clumpy, implying that the detected signal exhibits small-scale variations as a function of azimuth. Nevertheless, based on the above described procedure, we can assess that the global signal is statistically significant for all three features.

We then perform an additional test by calculating the statistical significance of the observed features using the wavelet transformation (Figure \ref{Fig:overview}C). To this end, we resort to the Multiresolution Analysis software (see Appendix\,\ref{appendix_wavelet}) and use it to compare the wavelet coefficients obtained from the data at each scale to the analytical distribution of coefficients that we would have obtained if our image had been flat (i.e. no signal) and affected by Poisson noise. We find that the identified spiral features in the UMS are more significant than 3$\sigma$, consistent with the results presented above.






\begin{figure}
   \centering
   \includegraphics[height=6.5cm]{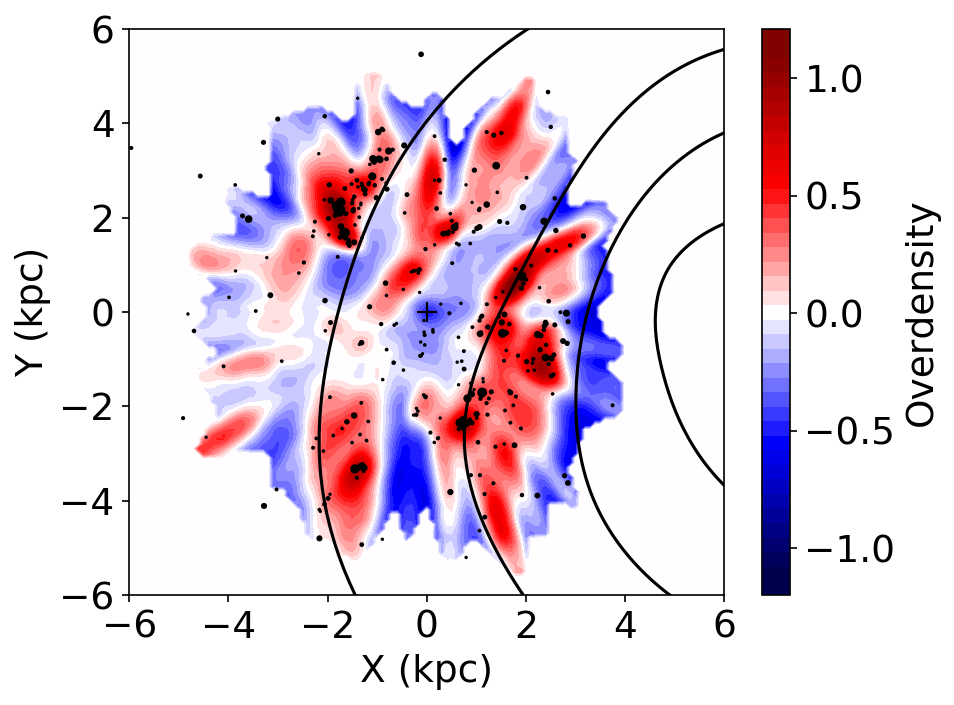}
   \caption{Same as Figure \ref{Fig:overview}B, but compared to the distribution of the young and instrinsically bright open clusters sample (see Section \ref{Sec:OCs}), shown by the black dots. The size of the dots is proportional to the number of cluster members brighter than absolute magnitude $M_G$> 0 (see text). Solid lines show the spiral arm model of \cite{Taylor:1993}, based on HII regions. }
              \label{Fig:OCs}%
\end{figure}

\begin{figure*}
   \centering
   \includegraphics[height=7.cm]{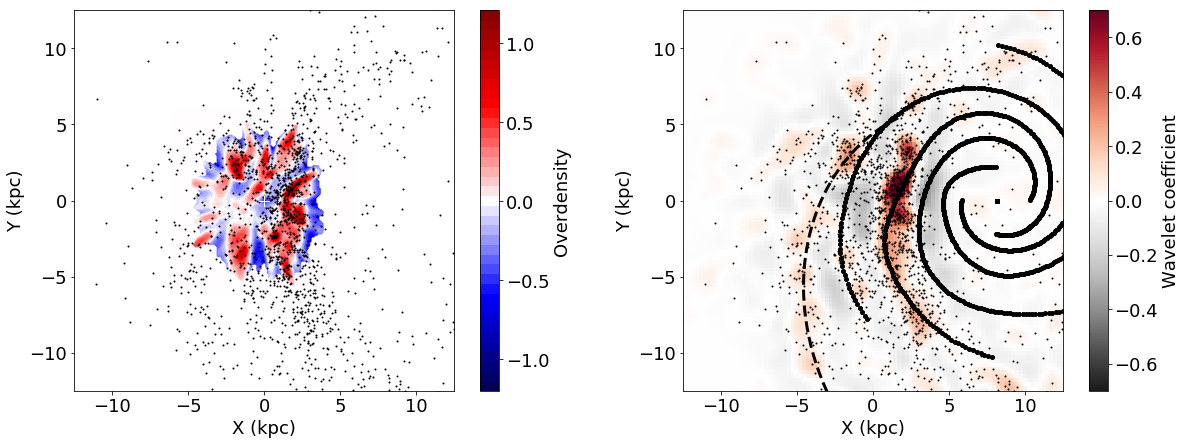}
   \caption{ \emph{Left panel:} Same as Figure \ref{Fig:overview}B, but on a larger scale, and compared to the distribution of the Cepheids sample (black dots). \emph{Right panel:} Wavelet transformation of the Cepheids sample, with oveplotted the positions of the single Cepheids (black dots), the L06 model for the Perseus arm (dashed curve) and the spiral arm model of \cite{Taylor:1993}, based on HII regions (solid lines).  }
              \label{Fig:Ceph}%
\end{figure*}


The spiral structure detected with two UMS samples can be then compared to the distribution of other spiral arm tracers. Figure \ref{Fig:OCs} shows the spatial distribution of the young ($<100$ Myr) and bright open clusters, compared to the UMS overdensity map. Given the young age of the selected OCs, we expect them to be good tracers of their birthplaces. 
As we can see from Figure \ref{Fig:OCs}, the OC's distribution is consistent with the overdense features found in the UMS sample, confirming that our UMS sample is indeed young, and therefore tracing
where star formation is more active. 

 Figure \ref{Fig:Ceph} (left panel) shows a comparison with the spatial distribution of the young ($<100$ Myr) Classical Cepheids. In contrast to the open clusters, within 5 kpc from the Sun there is not an obvious correspondence between the UMS overdensity map and the Cepheids' distribution. This may be due in part to our sample of Cepheids being on average older than the open clusters, so are more dispersed and not as constrained to the spiral arms. Indeed, while both samples were selected to have ages less than 100 Myrs, the open cluster sample has a mean age of 45 Myrs, while the Cepheids' mean age is 66Myrs. \citet{Skowron:2019} also show how the Cepheids diffuse with age. Also, while more numerous than the clusters, their individual distances (and ages) are less certain. 
 It is for these reasons that, we believe, the Local Arm is not clearly evident in the Cepheid sample.
 
 The advantage of the Cepheids is that they sample a much larger volume than the UMS stars and the open clusters, and we note that the UMS spiral arm segments might continue on a larger scale, as traced by the young Cepheids. In an effort to confirm this hypothesis we produced an overdensity map of the Cepheids, using the wavelet transform technique (Right panel of  Figure \ref{Fig:Ceph}). The small sample size limits the resolution that can be achieved (i.e. 1 kpc), so that the wavelet transform map cannot be done at the same scale as the UMS stars (i.e. 0.4 kpc), making the comparison between the two samples quite difficult. As in the left panel, the Local Arm is not evident near the Sun, though perhaps shows a weak feature in the third-quadrant. Nevertheless, the Sag-Car arm is clearly evident (right panel). If we assume the Perseus arm model from \citet{Levine:2006}, we can see that it is traced out to a much larger distance from the Sun compared to the UMS sample.
 However, by estimating the statistical significance of the wavelet transform (in the same way as previously done for the UMS stars), we find that the signal of the Cepheids along the Perseus arm is more significant than 3$\sigma$ only in few places, indicating a weaker signal compared to the one detected in the UMS sample, but nevertheless consistently tracing the proposed geometry of \citet{Levine:2006} for this arm (see discussion below).

The maps presented in this Section are publicly available at \url{https://github.com/epoggio/Spiral_arms_EDR3.git}.

\begin{figure}
   \centering
   \includegraphics[height=6.5cm]{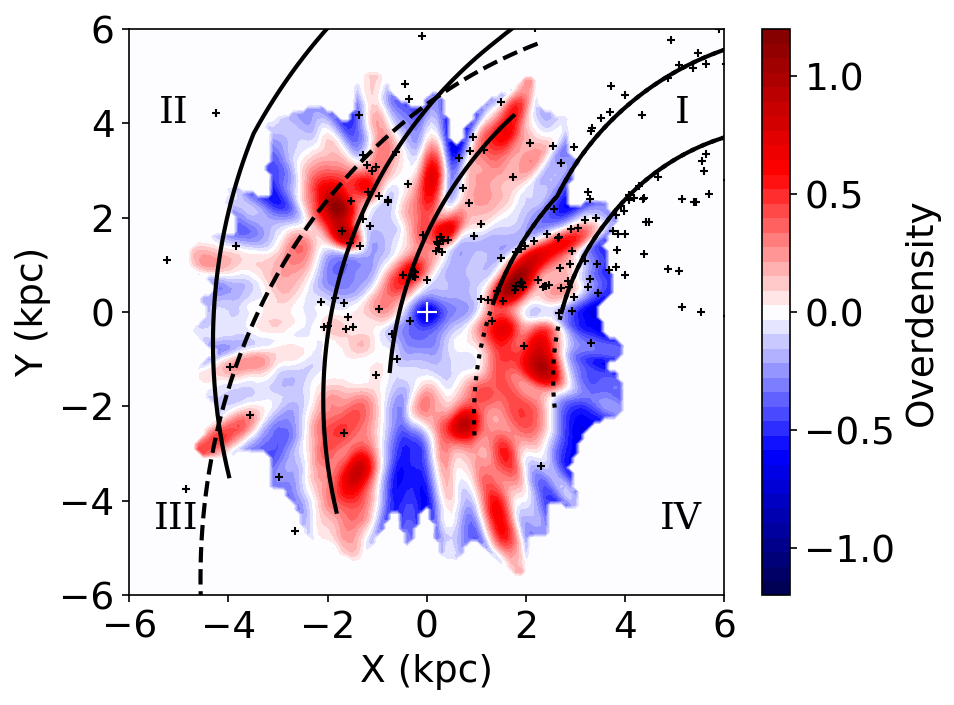}
   \caption{Comparison between the measured overdensity map presented in this work (Figure \ref{Fig:overview}B), the distribution of the maser sources (black dots) and spiral arm model (solid lines, from left to right: Outer, Perseus, Local, Sagittarius-Carina, Scutum arm) from \citet{Reid:2019}, and the Perseus arm from \citet{Levine:2006} (dashed line). Roman numerals show the I, II, III and IV Galactic quadrants. }
              \label{Fig:masers}%
\end{figure}


\section{Discussion} \label{Sec:DiscConcl}

It is worth comparing our overdensity maps, based on the distribution of young UMS stars, together with the open clusters and classical Cepheids, to previous models of the location of the spiral arms and other datasets avaiable in literature. By comparing our UMS overdensity map to the OB associations in the I and II quadrant first compiled by \citet{Morgan:1953} (see their Fig. 1), using updated distances from Benjamin (2021, priv. comm.), we found that there is excellent agreement (figure here not shown). 

In Figure \ref{Fig:OCs} we have overlayed the spiral arm model based on positions of HII regions \citep[][hereafter TC93]{Georgelin:1976,Taylor:1993}. The same model is also overlayed on the overdensity map of the Cepheids in Figure \ref{Fig:Ceph}, together with 
another model of the Perseus arm from \citet{Levine:2006} (hereafter L06), shown as a dashed line (arm n.2 in their paper). As mentioned in the introduction, this model is based on $21$cm emission from atomic hydrogen (HI), and covers the outer portions of the Galactic disk, which extends well outside the regions covered by our overdensity maps. 

Figure \ref{Fig:masers} shows a comparison between our overdensity map and the distribution of maser sources from \citet{Reid:2019}. We see that in the I and II galactic quadrants ($Y > 0$) there is reasonably good correspondence between the UMS overdensity and the maser positions (black crosses). 
Unfortunately, due to the current lack of maser data, a comparison in the III and IV quandrants is not possible. 
Figure \ref{Fig:masers} also shows the position of the spiral arms according to the model proposed by \citet{Reid:2019} (solid black curves, hereafter referred to as the R19 model), 
limited to the spatial range specified in their Table 2. For the two innermost arms (Sagittarius-Carina and Scutum), we show an extrapolation of their model (dotted lines), to explore the connection with our overdensity map. (For a more extensive extrapolation of their model, see their Fig. 1 and Fig. 2). 

As can be immediately seen, the UMS overdensity map shows a series of overdense regions that follow quite well the orientation of the Perseus arm as proposed by L06, with its rather steep pitch angle, rather than that of R19 or that of TC93 (Figure \ref{Fig:OCs}) based on HII regions. The distribution of the Cepheids, which extend significantly beyond the area of the disk covered by our overdensity map, also shows evidence of an outer structure consistent in orientation with that of the Perseus arm proposed by L06 (Figure \ref{Fig:Ceph}).

Another important difference between the R19 and TC93 models and the UMS overdensity in the III quadrant is the Local Arm. In our overdensity map this arm seems to continue into the III quadrant, well beyond the limited extent that has usually been proposed to-date for this arm. \citet{Vazquez:2008} and \citet{Xu:2021} also both suggest that the Local arm extends further than previously thought, though they propose significantly different geometries: \citet{Xu:2021} proposes that the Local Arm turns inward toward the Sag-Car arm, rather than into the III quadrant, while for \citet{Vazquez:2008} it continues in the direction of galactic longitude $240^\circ$ until it connects to the Perseus arm. For the R19 and TC93 model, the main overdensity of UMS stars in the III quadrant, centered on $(X,Y) = (-1.75, -3)$ kpc, would be assigned to the Perseus arm. Indeed, it has been remarked that toward the Galactic anticenter there seems to be a large gap in the Perseus arm \citep{Vazquez:2008,Bukowiecki:2011, CantatGaudin:2018, CantatGaudin:2019}. In our overdensity map, this gap is more naturally explained as simply being the interarm region between the Perseus arm, as traced by L06, and the Local arm extending into the III quadrant. 

Finally, the overdensities associated with the Sag-Car arm in the I and IV quadrants ($X > 0$) appear to be in good agreement with the TC93 model based on the HII regions, as well as the extrapolation of this arm in the R19 model. Toward the Galactic center we see a gap in the distribution of the UMS stars, clearly seen in the surface density (Fig. \ref{Fig:overview}A), which might be the interarm gap between the Sag-Car arm and the inner Scutum arm. This innermost arm is better seen in the overdensity map of the Z21 sample (see Fig. \ref{Fig:Z21_UMS_overdens_map}). However, the distance to this arm is closer to the Sun than either the R19 or TC93 models suggest. Alternatively, we are seeing another arm segment, possible a branch off the Sag-Car arm. 





\section{Conclusions} \label{Sec:concl}

In this contribution we have studied the spiral structure of the Galactic disk as traced by UMS stars, young open clusters, and classical Cepheids using the new parallaxes in \gedrthree. We mapped the overdensity of the UMS stars using two different methods and two different samples, all leading to similar results. Three arm segments are clearly discerned, corresponding to the Perseus arm, the Local Arm, and the Sag-Car arm, while a short segment of the Scutum arm towards the Galactic center is less evident, but more easily seen in the Z21 sample. Our resulting overdensity map for the UMS stars suggests that the geometry of the spiral arms in the III galactic quadrant differs significantly from those proposed by many current models, such as R19 and TC93, which are to a large degree based on an extrapolation of the arms as traced from sources in the I and II quadrants. 

From the distribution of the UMS sample we find that the Perseus arm follows the geometry proposed by \citet{Levine:2006}, based on HI radio data, and the distribution of Cepheids in the III quadrant is not inconsistent with this hypothesis. In addition, we see that the Local Arm in the UMS stars extends into the III quadrant at least 4 kpc past the Sun's position, giving it a full length of at least 8 kpc. This is still not enough to consider it a major arm of the Milky Way, but certainly means that it is an important spiral feature and more than a minor spur. Finally, the Scutum arm appears to be closer to the Sun than previously thought. 

The morphology of the spiral arms in the Galaxy 
is intrinsically connected to 
the mechanism responsible for their origin, which is still uncertain \citep{Toomre:1977,Athanassoula:1984,Binney:2008,Shu:2016}. 
Therefore, the large-scale mapping of the spiral arms potentially contains important clues on their nature, possibly helping to learn more about whether they have been triggered by the passage of a satellite \citep[][]{Purcell:2011,Laporte:2018,Bland-Hawthorn:2021} 
and/or other mechanisms which might be at play \citep[e.g.][]{Donghia:2013}.
 
In any case, to discern the dynamical nature of the spiral arms of the Milky Way, kinematic information will also be needed, and in particular the in-plane systematic motions. Using stellar kinematics, \citet{Baba:2018} found evidence that the Perseus arm is in the disruption phase of a transient arm. Recently, \citet{Pantaleoni:2021} identified a kinematically distinct structure, dubbed the Cepheus spur, extending from the Local arm towards the Perseus arm. This might be the same branch or spur off the Local Arm first noted by \citet{Morgan:1953}, and later mapped by \citet{Humphreys1970}. \citet{Pantaleoni:2021} suggest that this feature might be related to the Radcliffe wave \citep{Alves:2020}, a coherent gaseous structure in the solar neighbourhood, extending for 2.7 kpc in length. In the future we can look forward to having line-of-sight spectroscopic velocities in upcoming \gdrthree\ release to compliment the \gedrthree\ astrometry. 

While our distances are based on \gedrthree\ astrometry, our UMS source lists are largely based on photometry from 2MASS. In the future we can expect a significant increase in the volume of the disk that will be sampled by \gaia, thanks to the astrophysical parameters that are anticipated to be released in \gdrthree, based on \gaia\ spectrophotometry and astrometry for stars as faint as $G=18$. The ability to select young populations to fainter magnitudes will push to even larger distances our horizon out to which accurate Galactic cartography can be achieved for young luminous stars.

\begin{acknowledgements} 

We thank Robert Benjamin and Elena D'Onghia for fruitful discussions. This work has made use of data from the European Space Agency (ESA)
mission Gaia (https://www.cosmos.esa.int/gaia), processed by the Gaia Data Processing and Analysis Consortium (DPAC, https://www.cosmos.
esa.int/web/gaia/dpac/consortium). Funding for the DPAC has been provided by national institutions, in particular the institutions participating in the Gaia Multilateral Agreement.

This publication makes use of data products from the Two Micron All Sky Survey, which is a joint project of the University of Massachusetts and the Infrared Processing and Analysis Center/California Institute of Technology, funded by the National Aeronautics and Space Administration and the National Science Foundation.

EP acknowledges support by the Centre national d'études spatiales (CNES). TCG, FF, and MRG acknowledge support by the Spanish Ministry of Science, Innovation and University (MICIU/FEDER, UE) through grants RTI2018-095076-B-C21 and the Institute of Cosmos Sciences University of Barcelona (ICCUB, Unidad de Excelencia ’Mar\'{\i}a de Maeztu’) through grant CEX2019-000918-M.
RB, YF, and AL acknowledge support by the BELgian federal Science Policy Office (BELSPO) through various PROgramme de Développement d’Expériences scientifiques (PRODEX) grants. RA \& MF  acknowledge the support from the DLR (German space agency) via grant 50 QG 1403. 
PR acknowledges support by the Agence Nationale de la Recherche (ANR project SEGAL ANR-19-CE31-0017), funding from the project ANR-18-CE31-0006 and from the European Research Council (ERC grant agreement No. 834148). LC acknowledges support by the ANID/FONDECYT Regular Project 1210992.

RD, GC, VR and TM acknowledge the support of the Italian Space Agency (ASI) for their continuing support through contract 2018-24-HH.0 to the National Institute for Astrophysics (INAF)

\end{acknowledgements}

\bibliographystyle{aa}
\bibliography{example} 

\begin{appendix} 

\section{New calibration of Cepheid PW relation} \label{appendix_PWrelation}

To recalibrate the PW relations for the classical Cepheids, we made an additional refined selection of the Cepheid sample, retaining only objects with the most secure subclassification in modes from the literature (Ripepi et al. in preparation) and \texttt{ruwe} parameter $<1.4$\footnote{Section 14.1.2 of "\gaia\ Data Release 2 Documentation release 1.2"; https://gea.esac.esa.int/archive/documentation/GDR2/}, remaining in the end with a sample of well characterized 852 and 396 DCEP\_F and DCEP\_1O, respectively. Then, we corrected the zero point offset of the parallaxes following the indications by \citet{Lindegren2020}. However, several parametrizations are available in literature. \citet{Zinn:2021} found an offset of 15 $\mu$as for sources with G<10.8, which may lead up to 10 \% systematics in our Cepheids sample. By testing such an impact, we found that it doesn't affect our conclusions on the spatial distribution of the Cepheids sample. Finally, we fitted the PW relation to the data in the form $W=a+b (\log_{10} P-P_0)$, where $W$ is the absolute Wesenheit magnitude, $P$ the period of each DCEP and $P_0$ is a pivoting period equal to 1.0 d and 0.3 d for DCEP\_F and DCEP\_1O, respectively. Different relations were searched for F and 1O mode DCEPs. 
To estimate the parameters of the PW relations, we adopted the formalism by \citet{Riess2018} 
that allows us to use the parallaxes in a linear way, keeping the symmetry of their uncertainty, and avoiding biases from the cuts in parallax values. According to the distance modulus definition, we can define a photometric parallax (in mas)
\begin{equation}
\varpi_{phot}=10^{-0.2(w-W-10)}\, ,
\end{equation}
where $W$ is the absolute Wesenheit magnitude found from the PW relation and $w$ is defined in Section \ref{Sec:Cepheids}.
Indicating with $\varpi_{EDR3}$ the zero-point corrected parallax from EDR3, we minimize the following quantity:
\begin{equation}
\chi^2=\sum \frac{(\varpi_{EDR3}-\varpi_{phot})^2}{\sigma^2}    
\end{equation}
where $\sigma$ is calculated by summing in quadrature the error on EDR3 parallaxes, the uncertainty of 0.01 mas on the parallax correction \citep{Lindegren2020}, the intrinsic dispersion of the PW relations \citep[taken from][]{Ripepi:2019}, and the photometric uncertainty on the apparent $w$, which we assumed conservatively fixed and equal to 0.03 mag (more details on the whole procedure can be found in Ripepi et al. in preparation). 
Note that we did not consider any dependence on metallicity, as this information is missing for the large majority of the DCEPs considered here. Finally we arrive at the following PW relations:
\begin{eqnarray}
    W_F=(-6.015\pm0.005)-(3.317\pm0.028) (\log_{10} P_F-1.0) \\
    W_{1O}=(-4.170 \pm 0.005)-(3.624\pm0.017) (\log_{10} P_{1O}-0.3)
\end{eqnarray}
where the errors on the parameters are the formal output of the minimization routine\footnote{ \citep[Scipy.optimize.minimize python package][]{Virtanen2020}}. $P_F$ and $P_{10}$ are the periods of the F and 1O mode pulsators, respectively. The DCEP\_F PW relation is very similar to that derived by \citet{Ripepi:2019} based on DR2 parallaxes, but with much smaller errors due to both a larger number of objects and more precise parallaxes. On the contrary, the relation for DCEP\_10 is  discrepant with respect to that from DR2, but in the latter case the errors were very large due to the small statistics (the present DCEP\_1O sample is almost three times that used by \citealt{Ripepi:2019}). 


\section{Adopted techniques} \label{appendix_method}

\begin{figure}[htb]

   \centering
   \includegraphics[width=0.5\textwidth]{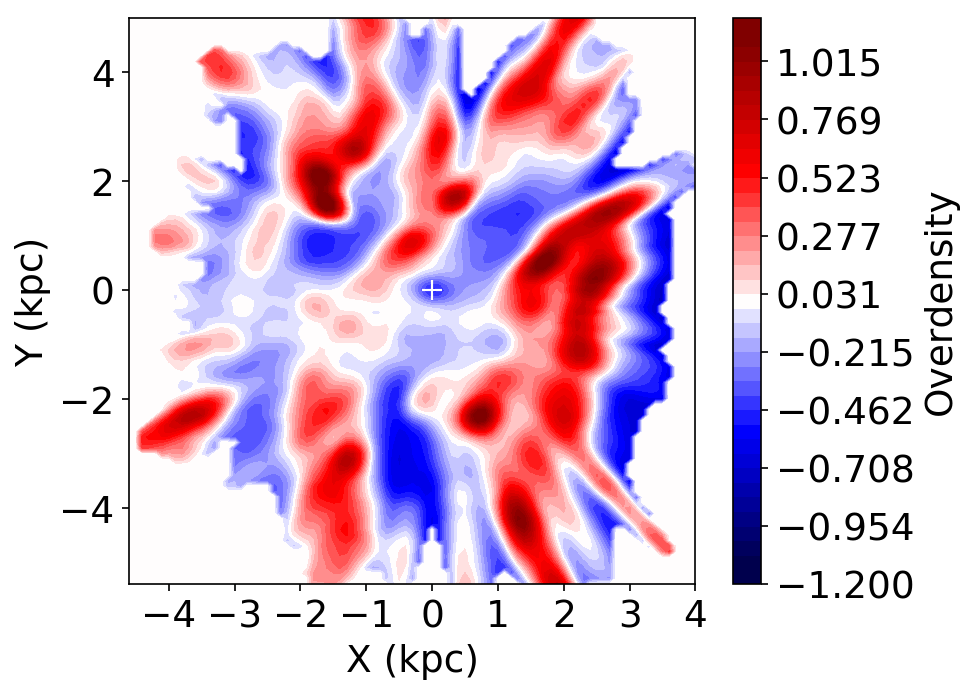}
   \caption{ Same as Figure \ref{Fig:overview}B, but using the Z21 UMS sample.   }
              \label{Fig:Z21_UMS_overdens_map}%
\end{figure}
\begin{figure}[hbt]

   \centering
   \includegraphics[width=0.45\textwidth]{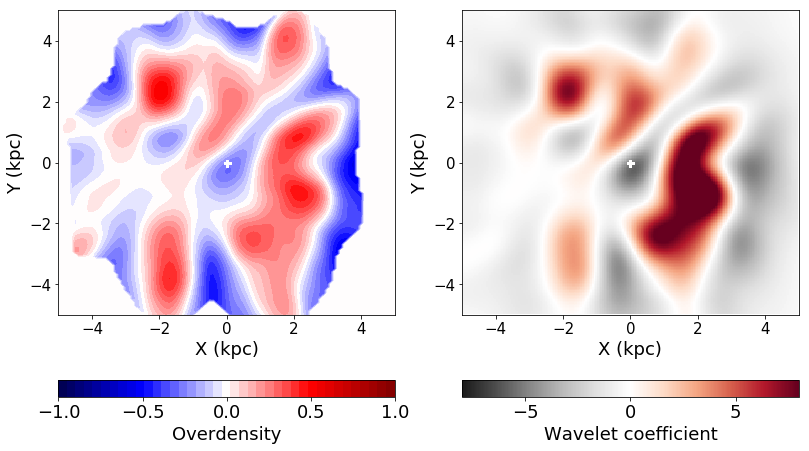}
   \caption{ Same as Figure \ref{Fig:overview}B (left panel) and C (right panel), but using a larger scale length $h=0.8$ kpc, which corresponds to scale=4 of the wavelet transform.   }
              \label{Fig:h08}%
\end{figure}

\subsection{Bivariate kernel density estimation}
\label{appendix_kernel}


For a given position (X,Y) in the Galactic plane, we calculate the local density $\Sigma(X,Y)$ through a bivariate kernel density estimator \citep[following Eq. 6.11 in][]{Feigelson:2012}, starting from the $(x_i, y_i)$-coordinates of the N stars in our UMS sample, where $i=1,...., N$:
\begin{equation}
 \Sigma(X,Y)  = \frac{1}{N \, h^2} \sum_{i=1}^{N} \Bigg[  K\bigg( \frac{X-x_i}{h} \bigg) \,  K\bigg(\frac{Y-y_i}{h}\bigg) \Bigg] \quad,
 \label{loc_dens}
\end{equation}
where $K$ is the kernel function and $h$ is the kernel bandwidth. Given the symmetry of the problem under study, here we adopt the same kernel function and the same bandwidth for both X- and Y- coordinates. In this paper, we have used the following Epanechnikov kernel function
\begin{equation}
 K\bigg( \frac{X-x_i}{h} \bigg) = \frac{3}{4} \, \bigg( 1 - \bigg( \frac{X-x_i}{h} \bigg)^2 \bigg)
\end{equation}
for $ |(X-x_i)/h|<1$, and is zero outside (and similarly for the Y-coordinate). However, we have also tested other kernel functional forms, such as the triangle and Gaussian kernel \citep[see for example][]{Feigelson:2012}, to check that the choice of the kernel does not significantly influence the obtained maps. The Figures presented in the main text are based on a local density bandwidth $h=0.3$ kpc. Here we also show the results obtained with a local density bandwidth $h=0.8$ kpc (Figure \ref{Fig:h08}).

To estimate the mean density $\langle \, \Sigma (X,Y) \, \rangle$, we use the same approach adopted for the local density (Equation \ref{loc_dens}), but choose a larger bandwidth $h=2$ kpc. Afterwards, the local and mean density are combined, following Equation \ref{Eq:overdensity}, to obtain the stellar overdensity at a given position (X,Y).

Additionally, we perform a test, by replacing the above-described mean density with the raw average of the local densities calculated within 2 kpc from each point. The resulting maps are consistent with the ones obtained above. 

\subsection{Wavelet transformation}
\label{appendix_wavelet}

The 2D wavelet transformation (WT, \citealt{Starck2002}) allows us to decompose an image into a set of layers, each one preserving only the structures with a characteristic size comparable to the wavelength of said layer (or scale). In terms of Fourier transformations, each layer of the WT can be roughly understood as the result of convolving the input with a pass-band that filters out those features in the original image that do not have the correct frequency.

There are many types of WT but, for the purpose of this work, we used the stationary wavelet transformation, also known as \emph{à trous} algorithm, in its redundant form and using the B-spline mother wavelet. We have computed the WT using the Multiresolution Analysis software \citep{Starck1998}, which has already been applied successfully to other astrophysical problems \citep[for more details on the WT transform see, e.g.][]{Ramos2018,Antoja2020,Ramos:2021}. Equation \ref{Eq:wt} shows the mathematical expression of the decomposition

\begin{equation}\label{Eq:wt}
    I(x,y) = c_p(x,y) + \sum_{j=1}^{p}\omega_j(x,y),
\end{equation}

where $I$ represents the original image, $c_p$ is an extremely smoothed version of it, and $\omega_j$ are the arrays of wavelet coefficients at each layer $j\in[0,p]$.

The advantatge of using this implementation of the WT in front of others is that the number of pixels are preserved and that each scale has zero mean. As a consequence of the latter, positive coefficients are directly related to overdensities and negatives ones, to underdensities, which is why panels A and B of Fig.~\ref{Fig:h08} appear visually similar. 


\subsection{Uncertainties}
\label{appendix_uncertainties}


Since the overdensity $\Delta_{\Sigma} (X,Y)$ is a derived quantity (Equation \ref{Eq:overdensity}), we calculate its uncertainty by first deriving the ones on the local density $\Sigma(X,Y)$ and the mean density $\langle \, \Sigma (X,Y) \, \rangle$. To this end, for a given position $(X,Y)$, we generate 100 bootstrap resamples using the stars within 0.3 and 2 kpc, respectively, for the local and mean density. Using the resamples, we calculate the bootstrap standard deviation $\sigma_{\Sigma(X,Y)}$ and $\sigma_{\langle \, \Sigma (X,Y) \, \rangle}$. The uncertainty on the overdensity $\Delta_{\Sigma} (X,Y)$ is then derived via error propagation:

\begin{equation}
\centering
\begin{split}
  \sigma_{\Delta_{\Sigma}(X,Y)} = \sqrt{ 
  \Biggl( \frac{\partial \Delta_{\Sigma}(X,Y)}{\partial \Sigma(X,Y)} \sigma_{\Sigma(X,Y)} \Biggr)^2 + 
  \Biggl( \frac{\partial \Delta_{\Sigma}(X,Y)}{\partial \langle \, \Sigma (X,Y) \, \rangle} \sigma_{\langle \, \Sigma (X,Y) \, \rangle} \Biggr)^2
  } \\
  = \sqrt{ 
  \Biggl( \frac{\sigma_{\Sigma(X,Y)} }{\langle \, \Sigma (X,Y) \, \rangle} \Biggr)^2 + 
  \Biggl( \frac{ \Sigma (X,Y)}{\langle \, \Sigma (X,Y) \, \rangle^2} \sigma_{\langle \, \Sigma (X,Y) \, \rangle} \Biggr)^2
  } \quad .
\end{split}
\end{equation}

\end{appendix}

\end{document}